\documentclass[aps,prb,twocolumn,superscriptaddress]{revtex4}

\usepackage{amsmath}
\usepackage{amssymb}
\usepackage{graphicx}

\begin{document}


\title[Exact GS of 1D LR RFIM]
{Exact ground states of one-dimensional long-range random-field
 Ising magnets}


\author{Timo Dewenter}
\email[]{timo.dewenter@uni-oldenburg.de}
\author{Alexander K. Hartmann}
\affiliation {Institut f\"ur Physik, Carl von Ossietzky Universit\"at 
Oldenburg, D-26111 Oldenburg, Germany}

\date{\today}
\begin{abstract}
We investigate the one-dimensional long-range 
random-field Ising magnet with Gaussian distribution of the random fields. 
In this model, a ferromagnetic bond between two spins is 
placed with a 
probability $p \sim r^{-1-\sigma}$, where $r$ is the distance between 
these spins and $\sigma$ is a parameter to control the effective dimension 
of the model. Exact ground states at zero temperature are calculated 
for system sizes up to $L = 2^{19}$ via graph theoretical algorithms for 
four different values of $\sigma \in \{0.25,0.4,0.5,1.0\}$
while varying the strength $h$ of the random fields. 
For each of these values several independent 
physical observables are calculated, i.e.\ magnetization, Binder
parameter, susceptibility and a specific-heat-like quantity.
The ferromagnet-paramagnet transitions at critical values $h_c(\sigma)$
as well as the corresponding critical exponents are obtained.
The results agree well with theory and interestingly we find 
for $\sigma = 1/2$ the data is compatible with a critical random-field 
strength $h_c > 0$.
\end{abstract}

\pacs{75.10.Nr, 75.40.−s, 75.50.Lk, 64.60.De}

\maketitle

\section{Introduction}
The critical behavior of spin systems with quenched disorder 
\cite{binder1986,mezard1987,young1998} is even today 
far from being well understood in contrast to pure models. Such a system 
with quenched disorder is 
the random-field Ising model (RFIM), where the spins interact 
ferromagnetically with each other and additionally
a quenched random field with strength $h$ acts locally
on the spins. In short-range models, 
it is known that the proposed equivalence \cite{Aharony76,Young77,Parisi79} 
of the critical behavior of 
a $d$-dimensional RFIM and a $(d-2)$-dimensional pure ferromagnet 
does not exist.A lower critical dimension of $d_c = 3$ for the RFIM 
resulting from the $d \rightarrow (d-2)$-rule was shown to be wrong 
\cite{Imbrie84}. The correct value of $d_c = 2$ was found 
by Imry and Ma \cite{Ma75} using their famous domain-wall argument
and later proven mathematically by Bricmont and Kupiainen
\cite{bricmont1987}.

A generalization of the short-range model are 
random-field Ising magnets with long-range interactions 
$J(r) \sim r^{-d-\sigma}$, the interaction strength $J$ decays like a 
power-law in the distance $r$. The exponent $\sigma$ allows the 
tuning of the effective dimensionality of the model,
allowing also for non-integer dimensions. Similar long-range
spin glass models, i.e.\, with bond disorder, have been studied recently
quite intensively for the case
of the fully connected model 
\cite{Katzgraber03,katzgraber05,1dchain_ultra2009,Leuzzi99}
as well as for the diluted case
\cite{leuzzi08,Leuzzi09,Katzgraber09,larson2013}. 
For the random-field Ising model, it  turned out that 
the proposed $d \rightarrow (d-\sigma)$ equivalence \cite{Grinstein76}, 
which is analogous to the 
$d \rightarrow (d-2)$-rule for short-range models, 
is wrong at higher orders of the pertubative expansion 
\cite{Young77,Bray86}. 
However, when one considers also long-range correlated random 
fields the situation is more interesting. \cite{Baczyk13}
A related 
model is the ferromagnetic hierarchical spin model introduced by Dyson 
\cite{Dyson69}, where the interaction strength decays exponentially with 
the level of the hierarchy. This model is solvable with exact 
renormalization and the hierarchical couplings are equivalent to 
long-range power-law couplings in real space. Because of this 
equivalence, the 
critical behavior of the Dyson hierarchical model with random 
fields \cite{Rodgers88,Monthus11} is expected to be the same as for 
one-dimensional long-range models with power-law interactions.

Further analyses of the RFIM with long-range 
interactions with renormalization-group theory \cite{Weir87,Bray86} or with 
mathematical tools \cite{Aizenman90,Aizenman89,Aizenman90_Err,Cassandro09} 
have been performed.
The result \cite{Bray86,Weir87,Cassandro09,Monthus11,Leuzzi13} that 
the lower 
critical dimension in short-range models ($d_c=2$) corresponds to the 
critical value $\sigma_c = 1/2$ in long-range models is obtained by a 
scaling argument similar to the Imry-Ma argument. 
In this argument no long-range order exists for $\sigma > 1/2$, whereas 
for $\sigma < 1/2$ a phase transition at zero 
temperature should occur.
The mathematical proofs by Aizenman and Wehr
\cite{Aizenman90,Aizenman89,Aizenman90_Err} which investigate the 
existence of such a phase transition require 
\cite{Aizenman90,Aizenman90_Err}
\begin{equation}
	|J_{x,y}| \leq c \cdot |x-y|^{-(3d/2 + \delta)}
	\label{J_ij}
\end{equation}
for the long-range interaction between spin $x$ and $y$, where $c$ is a 
constant and $\delta > 0$. Please note that the $\delta$ in 
Eq.\ \eqref{J_ij} was added later in an erratum,\cite{Aizenman90_Err}
which was published after the original article.\cite{Aizenman89}
We interpret Eq.\ \eqref{J_ij} in the way 
that for $d=1$ the value $\sigma = 1/2$ is excluded in the proof, so a 
phase transition for this value of $\sigma$ seems possible. In the proof 
of Cassandro, Orlandi and Picco 
\cite{Cassandro09} $\sigma = 1/2$ is also not taken into account, which 
allows for the existence of a phase transition for $\sigma = 1/2$ at 
$h_c > 0$.

Here, 
we use a slightly different model, where the couplings are random and 
only present with a certain probability, but the 
interaction strength $J$ has a fixed value. A central question is to 
find out whether there is a finite-disorder 
phase transition for the model studied here 
at zero temperature for the borderline case $\sigma = 1/2$.
For comparison we also consider few other
selected values of $\sigma$. In parallel and independently
of our work, the same question was tackled via considering the Binder
parameter and few other observables.\cite{Leuzzi13} 
For the present work, we consider beyond this a full
set of independent physical quantities, also involving the susceptibility
and a specific-heat-like quantity, to study the 
disorder-driven phase transitions and to obtain complete sets of critical
exponents.

The outline of this article is the following: First, the model is 
described, second the procedure to obtain a ground state for a given 
realization of the disorder 
is briefly outlined and third the physical observables and 
their expected scaling behaviors are explained. Next, results for the 
four investigated values of $\sigma$ are presented. Last, a conclusion 
which includes a comparison of the results with scaling relations and an 
outlook is drawn.

\section{Model}
\begin{figure}[ht]
	\centering
	\includegraphics[width=0.16\textwidth]{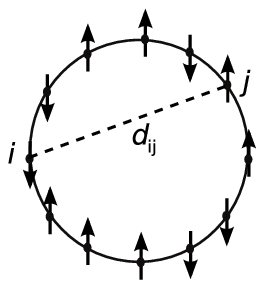}
	\includegraphics[width=0.2\textwidth]{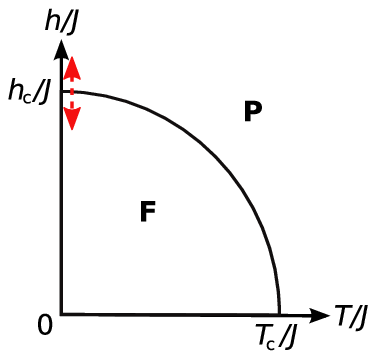}
	\caption{Left: One-dimensional spin-ring with $L=12$ Ising 
	spins. Right: Phase diagram of the Gaussian RFIM (corresponding 
	to Ref.\ \onlinecite{Hartmann02}), where ``F''
	denotes the ferromagnetic and ``P'' the paramagnetic phase, 
	both separated by the phase boundary. \label{ring_PD}}
\end{figure}

We study one-dimensional random-field Ising magnets with 
power-law diluted interactions, which are based on the 
one-dimensional long-range Ising chain.\cite{Ruelle68,Dyson69_2,Dyson71}
Instead of all-to-all coupling, 
where the interaction strength decays with a power law in the 
distance,\cite{Grinstein76,Bray86,Weir87} we use diluted 
interactions with fixed coupling strength, which recently have been 
used for spin glasses.\cite{leuzzi08, Katzgraber09} The Hamiltonian 
of the model used here is
\begin{equation*}
	\mathcal{H} = - J \sum_{i<j} \varepsilon_{ij} \; S_i \; 
			S_j - \sum_i (B_i + H) \; S_i,
\end{equation*}
where $J > 0$ (here we choose $J=1$) is the ferromagnetic 
coupling strength and the 
$S_i = \pm 1$ are Ising spins distributed on a ring with circumference 
$L$ (cf.\ Fig.\ \ref{ring_PD}). $B_i$ are the local random fields drawn 
from a Gaussian distribution with zero mean:
  \begin{equation*}
      p(B_i) = \frac{1}{\sqrt{2 \pi h^2}} \: \exp \left( 
		-\frac{B_i^2}{2 h^2} \right),
  \end{equation*}
where the width $h$ of the distribution controls the disorder strength.
The external homogeneous field $H$ is zero except for the determination 
of the susceptibility, where small fields are needed,
for technical reasons. The dilution 
matrix $\varepsilon_{ij}$ takes the value 1 if a bond is present 
between nodes $i$ and $j$ and 0 otherwise. A bond between non-nearest
neighbors on the ring exists with probability 
$p_{ij}$, where $p_{ij} \sim 1/d_{ij}^{1+\sigma}$ with 
$d_{ij} = (L/\pi) \sin (\pi |i-j|/L)$ (see Fig.\ \ref{ring_PD}) as 
geometric distance 
\cite{Katzgraber09, Katzgraber03} between two spins and $\sigma$ as 
parameter to control the effective dimensionality of the model. To avoid 
that $p_{ij} > 1$, one applies a short-distance cut-off 
\cite{Katzgraber09}, so that
\begin{equation*}
	p_{ij} = 1 - \exp \left( \frac{-A}{d_{ij}^{1+\sigma}} \right), 
	\quad z = \sum_{i=2}^{L-2} p_{iL}.
\end{equation*}
The constant $A$ is calculated numerically by fixing $z$, the average 
number of long-range bonds per node. As the nodes $1$ and $L-1$ are 
already neighbors of node $L$ on the ring, the sum to calculate $z$ 
starts at the next-nearest neighbor $2$.

The universality class of the model can be changed by varying $\sigma$. 
For $0 < \sigma < 1/3$ the critical exponents assume their 
mean-field (MF) values and for $1/3 < \sigma < 1/2$ 
the model is assumed to be in the non-MF region \cite{Monthus11}.
If $\sigma > 1/2$, one expects no phase transition,
\cite{Bray86,Weir87,Aizenman90,Aizenman89,Aizenman90_Err,Cassandro09,
Monthus11} 
i.e.\ the critical random-field strength $h_c = 0$ for $T=0$.

The MF values \cite{Bray86,Weir87,Monthus11} of the critical 
exponents are 
$\alpha = 0$, $\beta = 1/2$, $\gamma = 1$ and $\nu = 1/\sigma$. In the 
non-MF domain, i.e.\ $1/3 < \sigma < 1/2$ the correlation length exponent 
$\nu$ is not known exactly, so only the relations \cite{Monthus11}
\begin{equation}
 \frac{2-\alpha}{\nu} = 1-\sigma \quad \quad \frac{\beta}{\nu} = \frac 1 2 
 - \sigma \quad \quad \frac{\gamma}{\nu} = \sigma
 \label{crit_exp_NMF}
\end{equation}
are known analytically exact. But if, e.g., $\alpha$
is known ($\alpha=0$ seems plausible from the results presented below), 
the first relation in 
Eqs.\ \eqref{crit_exp_NMF} allows the determination of $\nu$
and thus of the other exponents.

Here, we focus on $\sigma = 0.25$, which 
belongs to the MF region, $\sigma = 0.4$ corresponding to the non-MF 
domain, $\sigma = 1/2$ right at the predicted border 
between non-MF region and the domain without a phase transition and 
$\sigma = 1$ from the $h_c = 0$ region.

\section{Obtaining Ground States}
The critical behavior of a Gaussian RFIM along the phase boundary is 
controlled by the zero-temperature fixed point \cite{Bray85}. Therefore, 
it is convenient to study the RFIM at $T=0$ and to alter the random 
field strength $h$ to cross the phase boundary (see arrow in Fig.\ 
\ref{ring_PD}). 
For the calculation of the exact ground state at $T=0$ for a given 
realisation the undirected graph is mapped to a directed network 
\cite{Picard75}. The maximum flow on this network is then calculated 
using a Push-and-Relabel algorithm \cite{Goldberg88}, whereof an 
efficient implementation exists in the LEDA-library \cite{LEDA}. These 
algorithms have a polynomial running time \cite{Ahuja93} and are faster 
than Monte-Carlo simulations (see e.g.\ Ref.\ \onlinecite{Rieger95}), 
because no 
equilibration time is needed and the ground state
is exact. After one has obtained the maximum flow, 
the directed network is mapped back to a ground-state spin configuration.

More details about the mapping to a directed network can be found in Ref.\
\onlinecite{Hartmann02}.

\section{Observables}
After obtaining the spin configuration of a ground state, we calculate 
physical quantities of interest. First, we fix $H=0$ and use $H > 0$ 
only for the calculation of the susceptibility. The average 
magnetization per spin is given by
\begin{equation}
	m = [|M|]_h = \left[ \left| \frac 1 N \sum_i S_i \right| 
							\right]_h,
	\label{mag_av}
\end{equation}
where $N \equiv L$ is the number of spins and $[\cdotp]_h$ denotes 
average over disorder. This averaging for fixed $h$ is performed over 
different realisations of graphs and random fields $\{B_i\}$, where for 
each configuration of long-range bonds one random-field realisation is 
used.

The Binder cumulant \cite{Binder81} is calculated via
\begin{equation}
	g(L,h) = \frac 1 2 \left( 3 - \frac{[M^4]_h}{[M^2]^2_h} \right),
	\label{binder_av}
\end{equation}
where in comparison to the original quantity the thermal average is 
omitted, because $T=0$ and the ground state 
is nondegenerate for a Gaussian RFIM.

To determine a specific-heat-like quantity \cite{Hartmann01} 
at $T=0$ we measure the bond energy
\begin{equation*}
	E_J = -\frac 1 N \sum_{i<j} \varepsilon_{ij} \; S_i \; S_j\,.
\end{equation*}
Now, we are able to 
differentiate $E_J$ numerically with respect to $h$ by calculating a 
finite central difference
\begin{equation}
	C \left( \frac{h_1 + h_2}{2} \right) = \frac{[E_J(h_1)]_h - 
					[E_J(h_2)]_h}{h_1-h_2},
	\label{spec_heat_av}
\end{equation}
which results in the specific-heat-like quantity $C$. The values $h_1$ 
and $h_2$ are two consecutive values of the random-field strength $h$, 
which have to be chosen appropriately.

The disconnected susceptibility is given by
\begin{equation}
	\chi_{\text{dis}} = L^d \; [M^2]_h,
\label{eq:chi:dis}
\end{equation}
in which $d = 1$ in our case.

For the determination of the susceptibility five different field 
strengths $H_n = n \cdot H_L$ with $n \in \{0,4\}$ of the 
homogeneous external field are applied to the system for each realisation 
and each value of $h$. A parabolic fit (for details see Ref.\ 
\onlinecite{Ahrens11}) 
to the datapoints yields the zero-field susceptibility
\begin{equation*}
	\chi = \left. \frac{\text{d} m}{\text{d} H} \right|_{H=0},
\end{equation*}
which is given by the slope of the parabola at $H=0$.

\subsection{Scaling in the non-mean-field region}
\label{Scaling_NMF}
For $\sigma > 1/3$, i.e.\ below the upper critical dimension the 
observables should scale close to the critical point $h_c$ like expected 
from finite-size scaling (FSS) theory (see e.g.\ Ref.\ 
\onlinecite{Yeomans92}).

The magnetization should scale like
\begin{equation*}
	m(h) = L^{-\beta/ \nu} \; \widetilde{m} ([h-h_c] \; L^{1/\nu}),
\end{equation*}
with some scaling function $\widetilde{m}$.

Close to the critical point, being a dimension-less quantity, 
the Binder parameter is assumed to have the 
following scaling behavior:
\begin{equation*}
	g(L,h) = \widetilde{g} ([h-h_c] \; L^{1/\nu}).
\end{equation*}

The scaling behavior of the singular part of the specific-heat-like 
quantity is
\begin{equation}
	C(h) = L^{\alpha/ \nu} \; \widetilde{C} ([h-h_c] \; L^{1/\nu}),
	\label{C_scal}
\end{equation}
and finite-size scaling predicts for the disconnected susceptibility
\begin{equation}
	\chi_{\text{dis}} (h) = L^{\overline{\gamma}/ \nu} \; 
	\widetilde{\chi}_{\text{dis}} ([h-h_c]\; L^{1/\nu}).
\label{eq:chi:dis:sclaing}
\end{equation}

The scaling behavior for the susceptibility is expected to be
\begin{equation*}
	\chi (h) = L^{\gamma / \nu} \; \widetilde{\chi} 
				([h-h_c] \; L^{1/\nu}).
\end{equation*}

\subsection{Scaling in the mean-field region}
For $0 < \sigma < 1/3$, i.e.\ above the upper critical dimension $d_u$ 
the usual finite-size scaling forms (cf.\ section \ref{Scaling_NMF}) are 
not valid (see e.g.\ 
Refs.\ \onlinecite{Luijten97,Jones05,Ahrens11,Monthus11}). 
At the critical point, the correlation length of the finite system 
is no longer proportional to the system size $L$, 
but behaves like \cite{Jones05,Ahrens11} $L^{d/d_u}$ and $L$ needs to 
be replaced \cite{Jones05} by $\ell = a_1 L^{d/d_u}$ in the FSS 
relations, where $a_1$ is a nonuniversal constant.
Therefore, the correlation length scaling exponent $\nu$ has to be 
replaced in the preceding section \ref{Scaling_NMF} to obtain scaling 
relations for the mean-field region by \cite{Botet82,Ahrens11}
\begin{equation}
	\nu^* = \frac{d_u}{d} \; \nu_{\text{MF}} = 3,
	\label{nu_star}
\end{equation}
where $d_u = 3 \sigma$, $d=1$ and $\nu_{\text{MF}} = 1/\sigma$ has been 
used. We therefore use $1/\nu^*=1/3$ instead of 
$1/\nu_{\text{MF}} = \sigma$ in the mean-field case $\sigma=1/4$ for our 
finite-size scaling analyses.

\subsection{Corrections to scaling at the lower and upper critical 
dimension}

Right at the upper critical dimension ($d_u = 4$) of the $\phi^4$-model, 
Br\'{e}zin \cite{Brezin82} showed that the correlation length 
$\xi \propto L (\log L)^{1/4}$. So, for $d=d_u$ logarithmic corrections
\cite{Jones05,Ahrens11} to 
scaling are expected and the lattice length $L$ has to be replaced by 
$\ell = a_2 L (\ln L)^{1/d_u}$.

Right at the lower critical dimension Leuzzi and Parisi \cite{Leuzzi13} 
recently 
proposed a logarithmic finite-size scaling. For the Binder parameter as 
well as for the two-point disconnected correlation function good data 
collapses for $\rho = 1.5$ (corresponding to $\sigma = 0.5$) and 
$(h/J)_c = 2.31(5)$ were achieved with logarithmic scaling.

In section \ref{sec_sigma05} we investigate the scaling behavior of some 
observables for $\sigma = 0.5$ to check whether an algebraic or 
logarithmic scaling appears.

\section{Results}
Next, we present the simulation results for the different values of 
$\sigma \in \{0.25,0.4,0.5,1\}$. System sizes from $L = 2^6 = 64$ up 
to $L = 2^{19} = 524288$ spins and $10^3$ to $10^6$ samples were used. 
All shown data points are averages over the given number of samples and 
the statistical errors result from the bootstrap resampling method 
\cite{Hartmann09}.
The average number of long-range bonds per node is fixed to $z = 6$.
For the determination of the susceptibility, the applied field stride 
$H_{L}$ of the homogeneous field is shown in Tab.\ \ref{small_field}.

\begin{figure}[ht]
	\centering
	\includegraphics[width=0.45\textwidth]
	{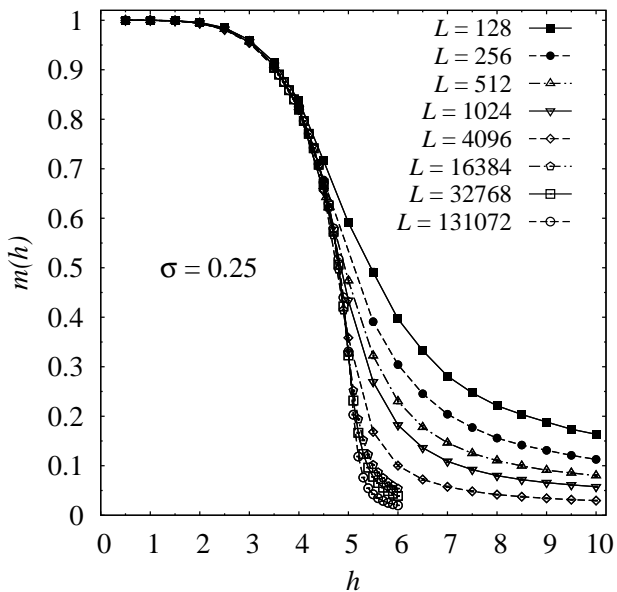}
	\caption{Average magnetization as a function of random-field 
	strength $h$ for different system sizes $L$ and $\sigma = 1/4$. 
	Data points are averaged over at least $10^3$ 
	samples and error bars result from 30 bootstrap samples. 
	Lines are guides to the eyes only. \label{mag}}
\end{figure}

\subsection{Mean-field region $\sigma = 0.25$}
Figure \ref{mag} shows the average magnetization per spin 
calculated by formula \eqref{mag_av} as a function of disorder 
strength $h$. For small $h$ the system is in the ferromagnetic ordered 
phase, where $m (h) \approx 1$ and for larger values of the 
random-field strength the system is in the paramagnetic phase, where 
$h \rightarrow 0$. With increasing $L$ the curves get steeper 
suggesting a phase transition at a critical value of $h_c \approx 5$.
\begin{figure}[htb]
	\centering
	\includegraphics[width=0.45\textwidth]
	{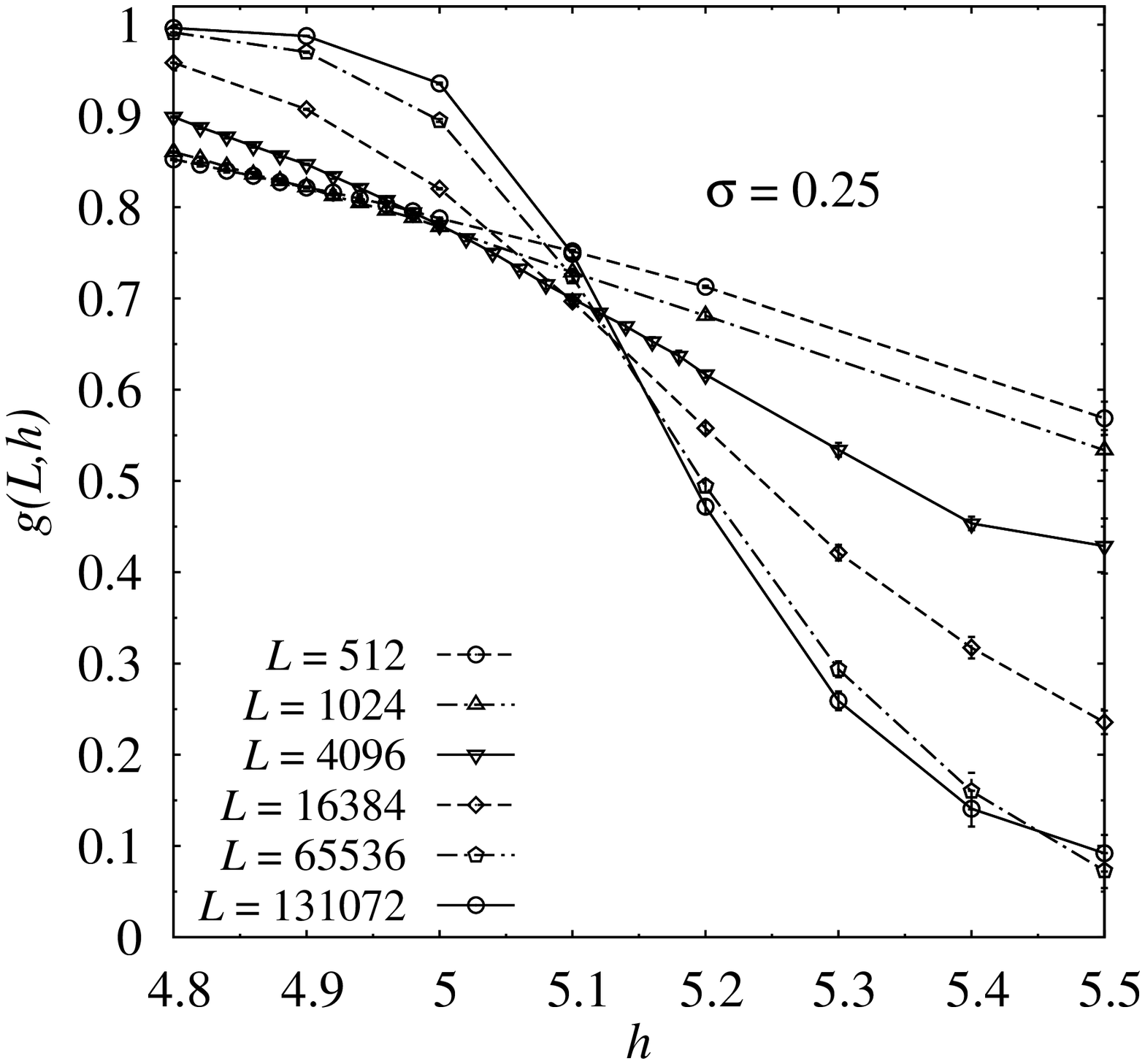}
	\caption{Binder parameter as a function of random-field 
	strength $h$ for different system sizes $L$ and $\sigma = 1/4$. 
 	Data points are averaged over at least $10^3$ 
	samples and error bars result from 30 bootstrap samples. 
	Lines are guides to the eyes only. \label{binder}}
\end{figure}

To determine this 
critical random-field strength more accurate, we calculate the 
Binder parameter, given in equation \eqref{binder_av}. Finite-size 
scaling theory predicts an intersection of the curves for the Binder 
cumulant for different system sizes at the critical point $h_c$. This can 
be seen in Fig.\ \ref{binder}, from which we estimate $h_c \approx 5.1$.
\begin{figure}[h]
	\centering
	\includegraphics[width=0.45\textwidth]
	{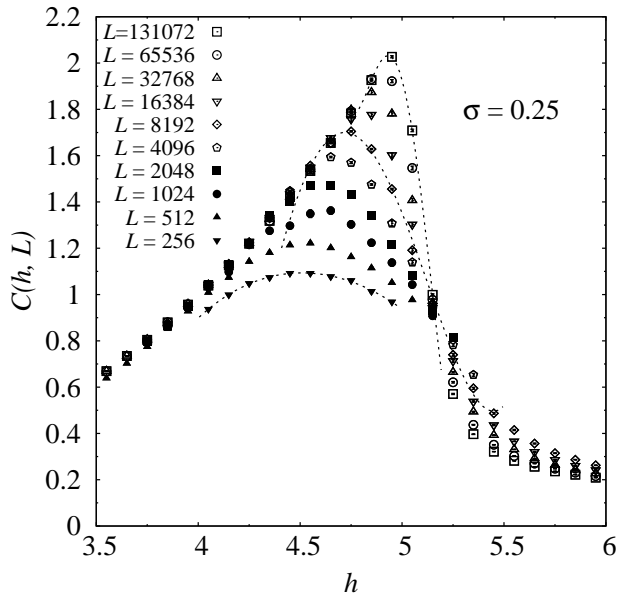}
	\caption{Specific-heat-like quantity $C$ averaged over at least 
	$10^4$ samples as a function of 
	random-field strength $h$ for different system sizes $L$ 
	and $\sigma = 1/4$. 
	Dashed lines are example fits for three system sizes with 
	fourth-order polynomials to obtain the maxima of $C$.
	\label{spec_heat}}
\end{figure}

Next, we investigate the specific-heat-like quantity $C$, where we 
choose $h$ values with distance $h_1-h_2 = 0.1$ in equation 
\eqref{spec_heat_av}. Figure \ref{spec_heat} shows the peaks of $C$ 
close to the critical point for different system sizes. One can 
observe that with increasing system size 
$L$ the peak height grows as well as the 
peak position shifts to larger values of $h$.
\begin{figure}[ht]
	\centering
	\includegraphics[width=0.45\textwidth]
	{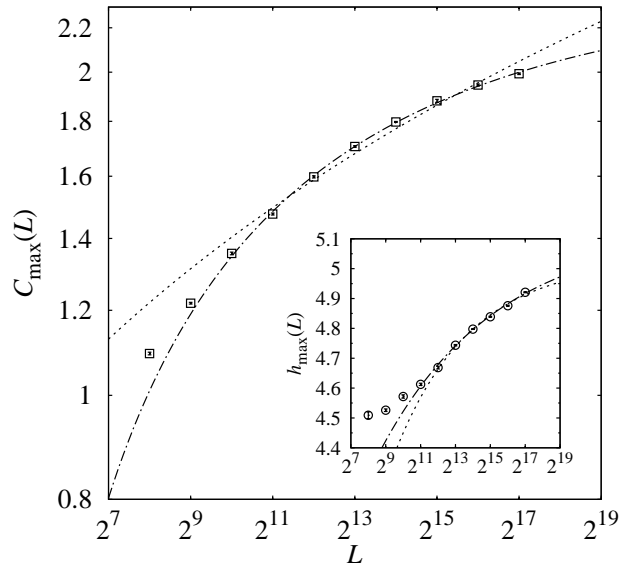}
	\caption{Double logarithmic plot of the 
	peak heights of the specific-heat-like quantity 
	$C$ as a function of system size $L$ for $\sigma = 1/4$. 
	Dotted line denotes logarithmic fit \eqref{logar} for $L>256$ 
	with parameters $a = 0.33(7)$, $b = 0.15(1)$ and 
	dash-dotted line is an 
	algebraic fit \eqref{alg} also for $L>256$ with parameters 
	$c = 2.49(7)$, $d = -1.52(9)$ and $k = -0.17(1)$. 
	Inset: Peak positions of $C$ as a 
	function of $L$. Dash-dotted line denotes a fit for $L>2048$ 
	by Eqs. \eqref{h_max}, where 
	$h_c = 5.13$, $a_2 = -2.7$ and $1/\nu^* = 0.215$. Dotted line 
	denotes same fit but fixed $1/\nu^* = 1/3$, resulting in 
	$h_c = 5.03$ and $a_2 = -5.82$.
	\label{C_peaks}}
\end{figure}

This impression is confirmed by Fig.\ \ref{C_peaks}. Apparently, 
both the 
peak heights and the peak positions behave like a power-law with added 
constant as a function of the number of spins $L$: In
fact we tested three different 
possible behaviors of the peak heights of the specific-heat-like 
quantity:
\begin{align}
	C_{\text{max}}^{\text{log}} (L) &= a + b \ln L , \label{logar} \\
	C_{\text{max}}^{\text{alg}} (L) &= c \cdot (1 + d \cdot L^k ), 
							\label{alg} \\
	C_{\text{max}}^{\text{corr alg}} (L) &= c_2 \; L^{\alpha/ \nu^*} 
		\cdot (1 + d_2 \cdot L^{k_2} ), \label{corr_alg}
\end{align}
a logarithmic divergence, an algebraic behavior and an algebraic 
function with a correction term.

All fits are least-squares fits with a reduced chisquare of
$\chi_{\text{red}}^2 = \sum_i^n [(y_i-f(x_i))/\Delta_i]^2/n_{\text{df}}$, 
where the degrees of freedom of the fit are 
$n_{\text{df}} = n-n_{\text{param}}$, which is the difference between the 
number of datapoints $n$ and 
the number of parameters $n_{\text{param}}$ in the fit-function $f$. The 
datapoints $(x_i, y_i \pm \Delta_i)$ have an error of $\Delta_i$.

The logarithmic fit yields a reduced chisquare of 
$\chi_{\text{red}}^2 \approx 200$ for system sizes $L>256$ and 
$\chi_{\text{red}}^2 \approx 118$ for $L>512$, 
which is quite bad. A better result is obtained with the algebraic fit 
where $\chi_{\text{red}}^2 = 6.9$ ($L>256$) or 
$\chi_{\text{red}}^2 =4.2$ for $L>512$, which is o.k. Because 
of these fits, a logarithmic divergence of the specific-heat-like 
quantity can be excluded.
The fit by equation \eqref{corr_alg} does not converge for values 
$\alpha / \nu^* > 0$, so that we conclude $\alpha / \nu^* = 0$.
\begin{figure}[h]
	\centering
	\includegraphics[width=0.45\textwidth]
	{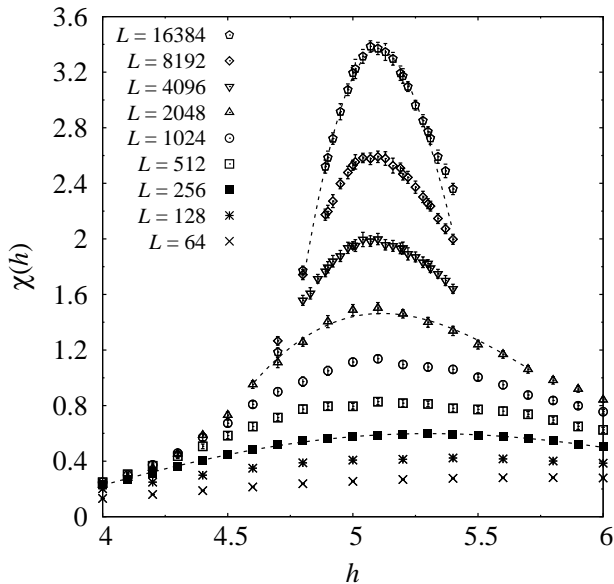}
	\caption{Susceptibility $\chi$ averaged over at least 
	$10^4$ samples with error bars resulting from 30 bootstrap 
	samples as a function of 
	random-field strength $h$ for different system sizes $L$ 
	and $\sigma = 1/4$. 
	Dashed lines are example fits for three system sizes with a 
	Gaussian and additional sigmoidal term to obtain the maxima of 
	$\chi$. Note that for $L=64$ and $L=128$ $\chi$-values up 
	to $h=8$ were used to determine the maxima, but are omitted here 
	for clarity of the plot. 
	\label{chi}}
\end{figure}

For the peak positions, fits of an algebraic function
\begin{equation}
	h_{\text{max}} (L) = h_c + a_2 \cdot L^{-1/\nu},
	\label{h_max}
\end{equation}
where $\nu = \nu^*$ should apply for the MF case $\sigma=1/4$.

\begin{table}[hb]
  \begin{tabular}{|c||p{7mm} c||p{7mm} c||p{7mm} c||p{7mm} c|}
	\hline
    & \multicolumn{2}{|c||}{$\sigma = 0.25$} & 
    \multicolumn{2}{|c||}{$\sigma = 0.4$} & 
    \multicolumn{2}{|c||}{$\sigma = 0.5$} & 
    \multicolumn{2}{|c|}{$\sigma = 1.0$} \\\hline
    $L$ & $H_L$ & $N_{\text{samp}}$ & $H_L$ & $N_{\text{samp}}$ & 
    $H_L$ & $N_{\text{samp}}$ & $H_L$ & $N_{\text{samp}}$ \\
     & & $/10^4$ & & $/10^4$ & & $/10^4$ & & $/10^4$ \\\hline
    64 & 0.300 & 100 & -- & -- & -- & -- 
	 & -- & -- \\
    128 & 0.065 & 10 & -- & -- & -- & --
	 & -- & -- \\
    256 & 0.050 & $5$  & 0.016 & $10$ & 0.0150 &  
	$5$ & 0.0250 & $5$ \\
    512 & 0.039 & $5$ & 0.011 & $10$ & 0.0110 & 
	$5$ & 0.0180 & $5$ \\
   1024 & 0.030 & $5$ & 0.008 & $5$ & 0.0075 & 
	$5$ & 0.0125 & $5$ \\
   2048 & 0.023 & $5$ & 0.006 & $5$ & 0.0050 & 
	$5$ & 0.0090 & $5$ \\
   4096 & 0.018 & $5$ & 0.004 & $5$ & 0.0038 & 
	$5$ & 0.0063 & $5$ \\
   8192 & 0.014 & $5$ & 0.003 & $5$ & 0.0027 & 
	$5$ & 0.0044 & $5$ \\
  16384 & 0.011 & $5$ & 0.002 & $1$ & 0.0019 & 
	$5$ & 0.0031 & $1$ \\

	\hline
  \end{tabular}
  \caption{System sizes $L$, smallest external fields $H_L$ 
  and number of samples $N_{\text{samp}}$ which are used to determine 
  the susceptibility for the given values of $\sigma$. 
  \label{small_field}}
\end{table}

Due to the change of curvature of the data, see
inset of Fig.\ \ref{C_peaks}, only system sizes $L > 2048$ 
were used for the fit. The fit by formula \eqref{h_max} gives 
$\chi_{\text{red}}^2 = 9.6$,
$h_c = 5.13 \pm 0.10$ and $1/\nu^* = 0.215 \pm 0.071$.
This value for $1/\nu^*$ is a bit off but still compatible 
within two error bars with the expected
$1/\nu^* = 1/3$. We also test, see 
the inset of Fig.\ \ref{C_peaks}, a fit by equation 
\eqref{h_max} for $L > 2048$ with fixed $1/\nu^* = 1/3$. 
It yields $\chi_{\text{red}}^2 = 13.6$ and the curves of both 
fits are quite close to each other, so $1/\nu^* = 1/3$ seems possible. 
Due to these results, i.e., 
strong finite-size corrections, the poor quality 
of the data for smaller system sizes and therefore the small amount 
of usable data points for the fits, the found value of $1/\nu^*$ is 
not included in the average given in Tab.\ \ref{crit_exp}.

Figure \ref{chi} shows the maxima of the zero-field susceptibility 
$\chi$, where the smallest external fields $H_L$, which were used to 
determine this quantity are given in 
Tab.\ \ref{small_field}.
\begin{figure}[ht]
	\centering
	\includegraphics[width=0.45\textwidth]
	{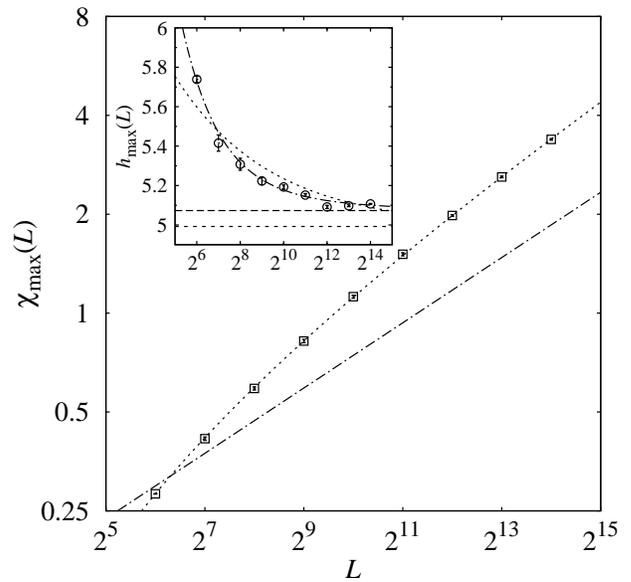}
	\caption{Double logarithmic plot of the 
	peak heights of the susceptibility $\chi$ as a 
	function of system size $L$ for $\sigma = 1/4$. 
	Dashed-dotted line denotes algebraic fit \eqref{chi_max} with 
	parameters $a_3 = 0.075$ and $\gamma / \nu^* = 0.33$ (fixed). 
	Dotted line is an algebraic fit with a correction term 
	\eqref{chi_max_corr} with parameters $a_4 = 0.170$, 
	$\gamma / \nu^* = 0.33$ (fixed), $d_3 = -1.330$ and $k_3 = -0.199$. 
	Inset: Peak positions of $\chi$ as a 
	function of $L$. Dotted line is a fit by Eq.\ \eqref{h_max} 
	with parameters $h_c = 4.993$, $a_2 = 2.39$, $1/\nu^* = 0.33$ (fixed)
	and dash-dotted line denotes a fit by Eq.\ \eqref{h_max_corr}, 
	where $h_{c2} = 5.073$, $a_5 = 0.6$, $1/\nu^* = 0.33$ (fixed), 
	$d_4 = 33$, $k_4 = -0.54$. 
	Horizontal lines are $h_{c2} = 5.073$ and $h_c = 4.993$, 
	respectively. 
	\label{chi_peaks}}
\end{figure}

It seems that the larger the system size $L$, the larger the peak 
height of $\chi$ and the (slightly) more the peak position
is at larger values of $h$.
This behavior is shown in Fig.\ \ref{chi_peaks}, where the maxima 
are expected to increase like
\begin{equation}
	\chi_{\text{max}} (L) = a_3 \cdot L^{\gamma / \nu^*}.
	\label{chi_max}
\end{equation}
A fit to the data with fixed value 
$\gamma/\nu^* = 0.33$ yields a
reduced chisquare of $\chi_{\text{red}}^2 \approx 2600$.
As visible from the double 
logarithmic plots in Fig.\ \ref{chi_peaks}, the data exhibits a clear
curvature, incompatible with a pure power law. When taking 
finite-size corrections into account and using
\begin{equation}
	\chi_{\text{max}} (L) = a_4 \cdot L^{\gamma / \nu^*} \cdot 
	(1 + d_3 \cdot L^{k_3})
	\label{chi_max_corr}
\end{equation}
again with fixed value $\gamma / \nu^* = 0.33$, this results
in $k_3 = -0.199 \pm 0.009$ and $\chi_{\text{red}}^2 = 0.31$.
This reduced chisquare value is much smaller than for a fit 
without corrections. Thus, the value of 
$\gamma / \nu^*$ seems to be appropriate.

The fits to the peak positions of the susceptibility are shown in the 
inset of Fig.\ 
\ref{chi_peaks}. A fit by formula \eqref{h_max} with fixed 
value $1/\nu^* = 0.33$ fit parameter yields 
$h_c = 4.993 \pm 0.023$ with
$\chi_{\text{red}}^2 = 33.6$.
\begin{figure}[ht]
	\centering
	\includegraphics[width=0.45\textwidth]
	{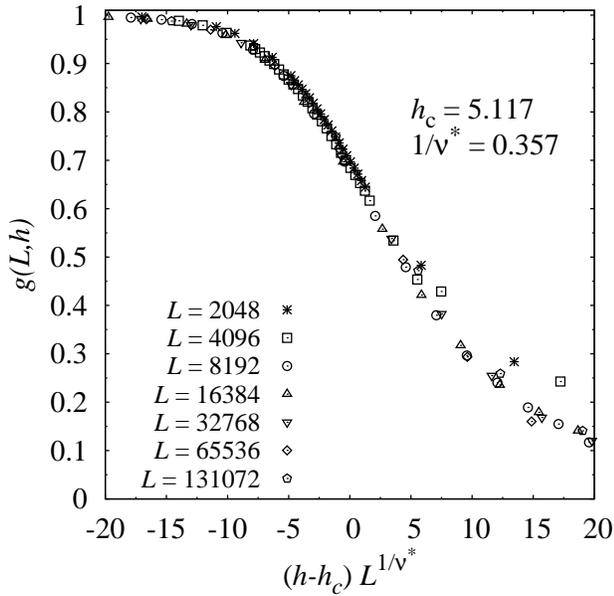}
	\caption{Data collapse of the Binder parameter for 
	$\sigma = 1/4$. Collapse was 
	performed for system sizes $L = 4096$ up to $L = 131072$. 
	\label{binder_coll}}
\end{figure}
A fit with correction term 
\begin{equation}
	h_{\text{max}} (L) = h_{c2} + a_5 \cdot L^{-1/\nu} \cdot 
	(1 + d_4 \cdot L^{k_4}),
	\label{h_max_corr}
\end{equation}
where here $\nu = \nu^*$ and fixed $1/\nu^* = 0.33$ gives 
$\chi_{\text{red}}^2 = 7.1$.
This value is smaller than for a fit without corrections, so we 
keep the chosen value $1/\nu^* = 0.33$. Further parameters of the 
fit by equation \eqref{h_max_corr} are $h_{c2} = 5.073 \pm 0.057$ and 
$k_4 = -0.54 \pm 0.72$.

\begin{figure}[hb]
	\centering
	\includegraphics[width=0.45\textwidth]
	{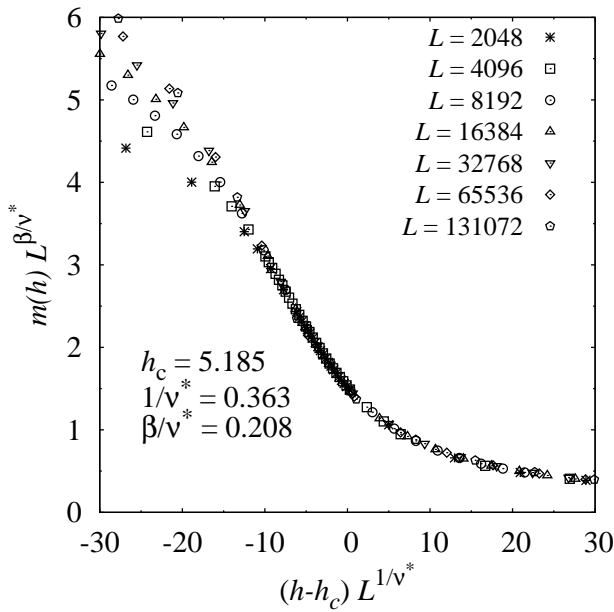}
	\caption{Data collapse of the magnetization for 
	$\sigma = 1/4$. Collapse was performed for system sizes from 
	$L=4096$ up to $L = 131072$. Smaller system size is shown for 
	comparison. 
	\label{mag_coll}}
\end{figure}

Next, we perform data collapses of the observables to
obtain estimates for the critical
exponents with another independent approach. For the determination 
of the best collapse we used a python script \cite{Melchert09}. 
Figure 
\ref{binder_coll} shows the collapse for the Binder cumulant 
with parameters $h_c = 5.117 \pm 0.005$ and $1/\nu^* = 0.357 \pm 0.027$. 
The value of $1/\nu^*$ is compatible with the expected value
$1/\nu^* = 1/3$ within the error bar. The 
quality of the collapse is very high below the critical point.
Above the critical point, only the two smallest system sizes exhibit
a notable deviation from a joint scaling curve, which can be attributed
to finite-size corrections to scaling.
\begin{figure}[ht]
	\centering
	\includegraphics[width=0.47\textwidth]
	{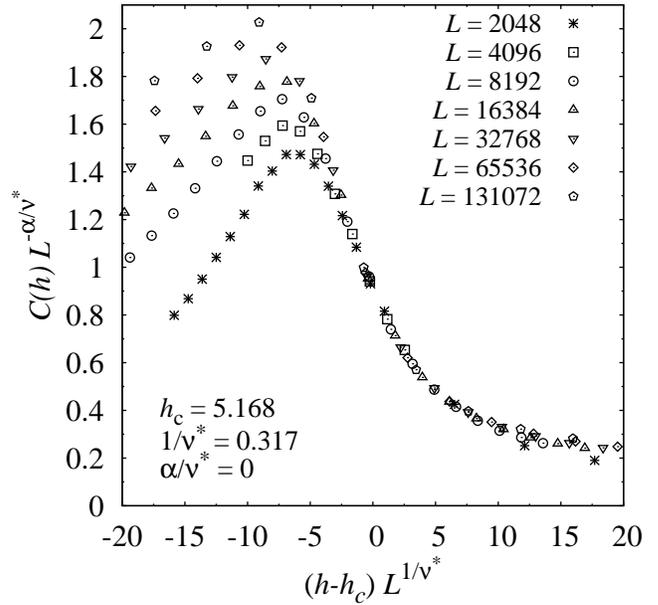}
	\caption{Data collapse of the specific-heat-like quantity for 
	$\sigma = 1/4$. System sizes from $L=2048$ up to $L=131072$ were 
	used for the collapse. Note that $\alpha = 0$ is fixed. 
	\label{C_coll}}
\end{figure}

The data collapse of the magnetization is presented in Fig.\ 
\ref{mag_coll}. The parameters of the collapse, which has
a high quality around the phase transition $h-h_c \approx 0$,
have the following values 
$h_c = 5.185 \pm 0.003$, $1/\nu^* = 0.363 \pm 0.020$ 
and $\beta/ \nu^* = 0.208 \pm 0.003$. This means 
$\beta = 0.573 \pm 0.041$, which is compatible within 
two standard error bars with the mean-field value 
$\beta = 1/2$.

The result of the data collapse for the specific-heat-like 
quantity is shown in Fig.\ \ref{C_coll}, where the important
parameters 
$h_c = 5.168 \pm 0.004$, $1/\nu^* = 0.317 \pm 0.010$ and 
$\alpha = 0$ (fixed) were used. Below the critical point, the collapse 
is poor, whereas around and above the critical point it is quite good.

\begin{figure}[h]
	\centering
	\includegraphics[width=0.45\textwidth]
	{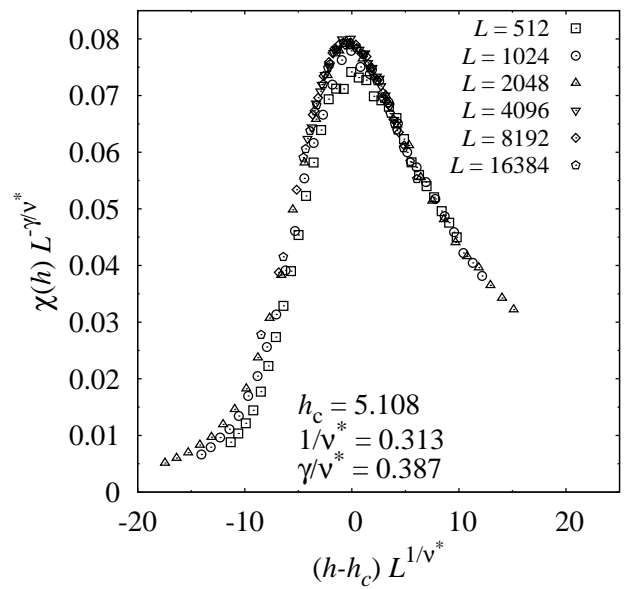}
	\caption{Data collapse of the susceptibility for 
	$\sigma = 1/4$. Collapse was 
	performed for system sizes $L = 2048$ up to $L = 16384$. Smaller 
	system sizes are included for comparison.
	\label{chi_coll}}
\end{figure}

The data collapse of the susceptibility is shown in Fig.\ 
\ref{chi_coll}. The important parameters
of the collapse are $h_c = 5.108 \pm 0.064$, 
$1/\nu^* = 0.313 \pm 0.064$ and $\gamma/\nu^* = 0.387 \pm 0.023$. 
The quality of the collapse is very good, except for smaller system 
sizes $L < 2048$, where deviations especially around the critical 
point occur.

Finally, the data collapse of the disconnected susceptibility
(not shown) for system sizes $L=2048$ up to $L=131072$ 
yields $h_c = 5.146 \pm 0.003$, 
$1/\nu^* = 0.342 \pm 0.001$ and $\bar{\gamma} / \nu^* = 0.666 \pm 0.005$.
This results in $\bar{\gamma} = 1.947 \pm 0.020$ which is compatible 
with the mean-field value $\bar{\gamma} = 2$ within three standard 
errors.

A summary of the results for all critical exponents is
shown in Table \ref{crit_exp}. We have obtained these values 
by averaging the results obtained by different methods, respectively.
The error bars are chosen such
that they include the values obtained by the different methods.
This should account for systematical errors, in particular corrections
to scaling. This results in all values being compatible with
the mean-field predictions.

\begin{table}[h]
  \begin{tabular}{|p{12.3mm} c | p{10mm}|p{10mm}|p{12mm}|p{4mm}|p{12mm}|p{12mm}|}
	\hline
    & & $h_c$ & $1/\nu$ & $\beta$ & $\alpha$ & $\gamma$ & 
    $\bar{\gamma}$ \\\hline
    $\sigma = 0.25$ & m & 5.13(6) & 0.34(6) & 0.62(13) & 0 & 1.06(29) & 1.98(39) \\
		    & t & 3.9-6.6 & 0.33 & 0.5 & 0 & 1 & 2 \\\hline
    $\sigma = 0.4$ & m & 4.5(2) & 0.30(6) & 0.27(8) & 0 & 1.50(54) & 2.74(54) \\
		    & t & 2.5 & 0.3 & 0.33 & 0 & 1.33 & 2.66 \\\hline
    $\sigma = 0.5$ & m & 3.7(2) & 0.25(9) & 0.06(3) & 0 & 2.00(85) & 3.8(13) \\
		    & t & -- & 0.25 & 0 & 0 & 2 & 4 \\\hline
    $\sigma = 1.0$ & m & 0 & 0.40(8) & 0 & 0 & 2.19(53) & 2.51(83) \\
		    & t & 0 & 0.5 & 0 & -- & 2 & 2 \\\hline
  \end{tabular}
  \caption{Results of the ground-state calculations for the 
  investigated values of $\sigma$ (line with ``m''). For comparison 
  the theoretical values 
  \cite{Bray86, Weir87, Grinstein83, Monthus11, Fisher01, Aharony83} are given, 
  where for $\sigma \in \{0.4,0.5\}$ $\alpha = 0$ was assumed to get 
  estimates of 
  the other exponents (cf.\ Eqs.\ \eqref{crit_exp_NMF}). 
  Note that for $\sigma=0.25$ the value for $1/\nu^*$ is given here.
  The value $\gamma = 2$ for $\sigma = 1$ was obtained 
  by calculating the susceptibility 
  $\chi = \lim_{H \rightarrow 0} \partial m/\partial H \sim h^{-2}$ using the 
  equilibrium magnetization from reference \onlinecite{Fisher01}. 
  Theoretical values for $\bar{\gamma}$ 
  were obtained by using the Schwartz-Soffer equation 
  \eqref{Schwartz_Soffer}, 
  except for $\sigma=1$, where trivially $\bar{\gamma}/\nu=d=1$ from 
  Eqs.\ \eqref{eq:chi:dis} and \eqref{eq:chi:dis:sclaing}.
  \label{crit_exp}
}
\end{table}

\subsection{Non-mean-field region $\sigma = 0.4$}

For the non-mean field region, we expect still
a clear phase transition but with different exponents. We have
performed
simulations and analyses in the same way as for $\sigma=0.25$. For
brevity, we omit most plots, since they look similar as for the
mean-field case.
\begin{figure}[ht]
	\centering
	\includegraphics[width=0.45\textwidth]
	{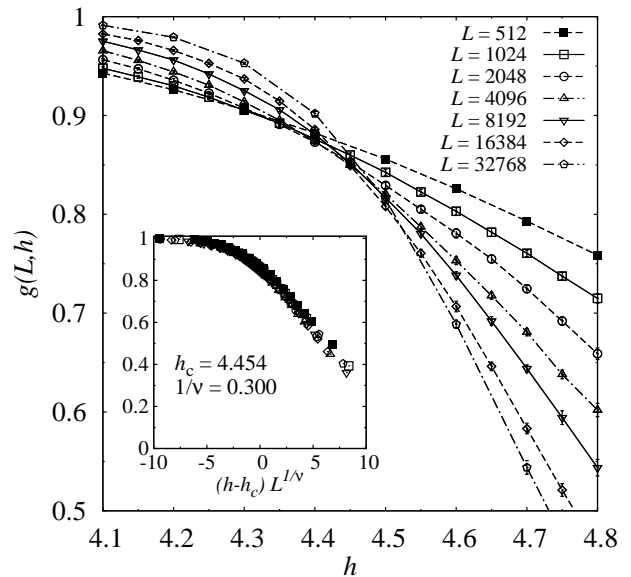}
	\caption{Binder parameter as a function of random-field 
	strength $h$ for different system sizes $L$ and $\sigma = 0.4$. 
	Lines are guides to the eyes only. 
	Inset: Data collapse of the Binder cumulant for $\sigma = 0.4$ 
	and system sizes $L = 2048$ up to $L = 32768$. Smaller sizes are 
	shown for comparison. 
	\label{binder_040}}
\end{figure}

As an example, Fig.\ \ref{binder_040} shows the Binder 
parameter as a function of the 
disorder strength $h$ for $\sigma = 0.4$. One can see an intersection 
of all curves close to $h_c \approx 4.45$ indicating a phase transition 
at this point. The inset presents the data collapse of the Binder 
cumulant which seems quite good, as the curves for the different system 
sizes fall onto one curve. The parameters for this collapse were 
$h_c = 4.454 \pm 0.015$ and $1/\nu = 0.300 \pm 0.058$.

We have obtained critical exponents for the other quantities
in the same way as discussed above. The results are summarized in
Tab.\ \ref{crit_exp}. In particular, $1/\nu = 0.30(6)$ agrees with 
$1/\nu = 0.316(9)$ from reference \onlinecite{Leuzzi13} (for $\rho=1.4$ in the 
cited paper).

\subsection{Borderline case $\sigma = 0.5$}
\label{sec_sigma05}

The value of $\sigma = 1/2$ was conjectured 
\cite{Bray86,Weir87,Cassandro09,Monthus11} to correspond
to the lower critical dimension. Thus, so for $\sigma >1/2$ one has 
$h_c = 0$.
Nevertheless, right at the critical value $\sigma=\sigma_c=1/2$, the
behavior could also correspond to $h_c>0$, as mathematical 
proofs \cite{Cassandro09,Aizenman90,Aizenman89,Aizenman90_Err} do not 
exclude the possibility of a phase transition for $\sigma=\sigma_c$.
We investigated this issue in the same way as for the
cases $\sigma<1/2$.

\begin{figure}[h]
	\centering
	\includegraphics[width=0.45\textwidth]
	{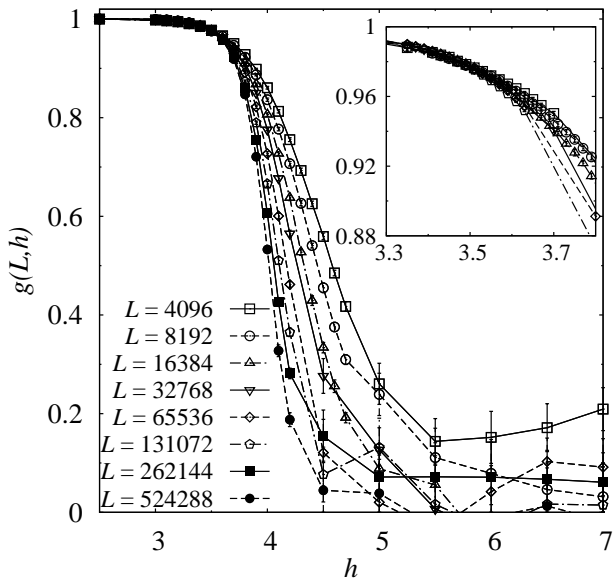}
	\caption{Binder cumulant as a function of the random field 
	strength for $\sigma = 1/2$.
	Inset: No clear intersection of the 
	curves for different system sizes $L$ can be determined. 
	Lines are guides to the eyes only.
	\label{binder_050}}
\end{figure}

The curves of the Binder cumulant (Fig.\ \ref{binder_050}) for 
different system sizes do not show a clear intersection. This
could be a hint towards $h_c=0$.

\begin{figure}[h]
	\centering
	\includegraphics[width=0.45\textwidth]
	{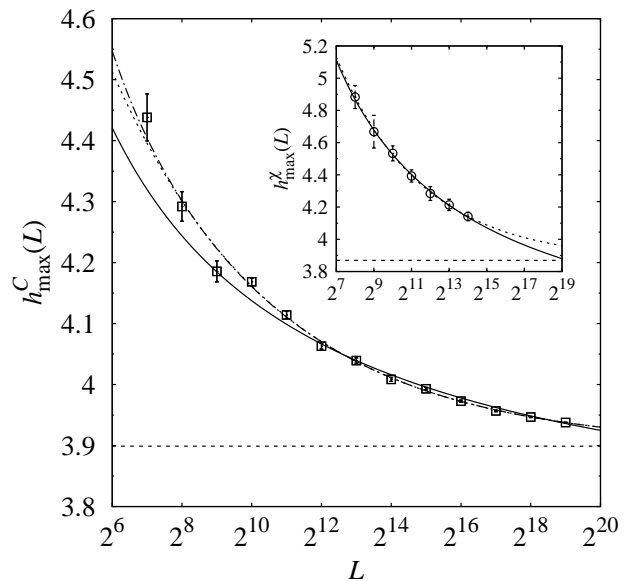}
	\caption{Peak positions of the specific-heat-like 
	quantity as a 
	function of system size for $\sigma = 1/2$. Dotted line is a fit 
	by Eq.\ \eqref{h_max} with parameters $h_c = 3.899$, $a_2 = 2.20$ 
	and $1/\nu = 0.307$. Dashed-dotted line is a fit by Eq.\ 
	\eqref{h_max_corr} with $h_{c2} = 3.898$, $a_5 = 2.13$, 
	$1/\nu = 0.302$, $d_4 = 11$ and $k_4 = -1$. 
	Horizontal line denotes $h_c = 3.899$. Solid line is a logarithmic 
	fit by Eq.\ \eqref{h_max_log} with $h_{c3} = 3.71$ and 
	$a_6 = 2.95$.
	Inset: Peak positions of the susceptibility as a function of 
	system size $L$. Dotted line is a fit by Eq.\ \eqref{h_max}, where 
	$h_c = 3.869$, $a_2 = 5.84$ and $1/\nu = 0.316$. Horizontal line 
	denotes $h_c = 3.869$. Solid line is a logarithmic fit by Eq.\ 
	\eqref{h_max_log} with $h_{c3} = 3.162$ and $a_6 = 9.46$.
	\label{hc_C_chi}}
\end{figure}

Thus, we studied the peak positions of the 
specific-heat-like quantity as shown in 
Fig.\ \ref{hc_C_chi}. When fitting a power law Eq.\ \eqref{h_max}
we obtained $h_c = 3.899 \pm 0.004$ and $1/\nu = 0.307 \pm 0.014$ 
with a quality of the fit of $\chi_{\text{red}}^2 = 1.5$.
This strongly indicates $h_c \approx 3.9 >0$. 
Note that we also fitted a power-law with correction term 
\eqref{h_max_corr}. The important parameters are 
$h_{c2} = 3.898 \pm 0.008$, $k_4 = -1 \pm 12$ and 
$1/\nu = 0.302 \pm 0.035$. The reduced chisquare is now 
$\chi_{\text{red}}^2 = 1.9$.
To check for logarithmic scaling \cite{Leuzzi13} another fit 
function was taken into account:
\begin{equation}
	h_{\text{max}} (L) = h_{c3} + \frac{a_6}{\ln L},
	\label{h_max_log}
\end{equation}
which leads to $\chi_{\text{red}}^2 = 8.4$ with the parameters 
$h_{c3} = 3.71 \pm 0.01$ and $a_6 = 2.95 \pm 0.12$.
Thus, a logarithmic scaling assumption seems less compatible with our 
results than a power-law behavior (with corrections).

Furthermore, we obtained the susceptibility and the corresponding
positions (and heights) of the peaks.
In the inset of Fig.\ 
\ref{hc_C_chi} the data for the peak positions of the susceptibility and 
fits are presented. The first one  by equation \eqref{h_max} 
yields a reduced chisquare of $\chi_{\text{red}}^2 = 0.03$ with 
$h_c = 3.869 \pm 0.033$ and an exponent $1/\nu = 0.316 \pm 0.021$.
The second fit by Eq.\ \eqref{h_max_log} yields with 
$h_{c3} = 3.162 \pm 0.023$ and $a_6 = 9.46 \pm 0.19$ to 
$\chi_{\text{red}}^2 = 0.06$, so both fits are compatible with our data.
And indeed, as the inset of Fig.\ \ref{hc_C_chi} shows both curves agree
very well in the range of the data points.

Thus, our results clearly support $h_c>0$ for $\sigma=0.5$. 
Although we cannot determine whether the finite-size scaling is of 
logarithmic or of power-law type, both suggest that $h_c > 0$.
Recent results 
which support our findings were provided by Ref.\ \onlinecite{Garel13}, 
where the Dyson hierarchical random-field model (cf.\ Ref.\ 
\onlinecite{Monthus11}) for 
$\sigma = 1/2$ was investigated numerically for system sizes  up to 
$L = 2^{21}$. These results strongly indicate
that the magnetization converges for system sizes $L \rightarrow \infty$ 
to one common curve at $h_c > 0$.
In reference \onlinecite{Leuzzi13}, Binder cumulants of a 
one-dimensional RFIM on a L\'{e}vy lattice are studied. Finite-size 
scaling analysis of the Binder parameter at the value 
$\sigma = 1/2$ (corresponding to $\rho=3/2$ in the cited paper)
yielded \cite{Leuzzi13} $(h/J)_c \approx 2.31(5) > 0$.

Finally note that also 
the data points of the magnetization (not shown) for 
various system sizes converge for $L \rightarrow \infty$ to one single 
curve with $h_c \approx 4.0 > 0$. The complete set
of resulting estimates for the critical exponents is again
shown in Tab.\ \ref{crit_exp}.

\subsection{Region without non-trivial phase transition $\sigma = 1.0$}

\begin{figure}[h]
	\centering
	\includegraphics[width=0.45\textwidth]
	{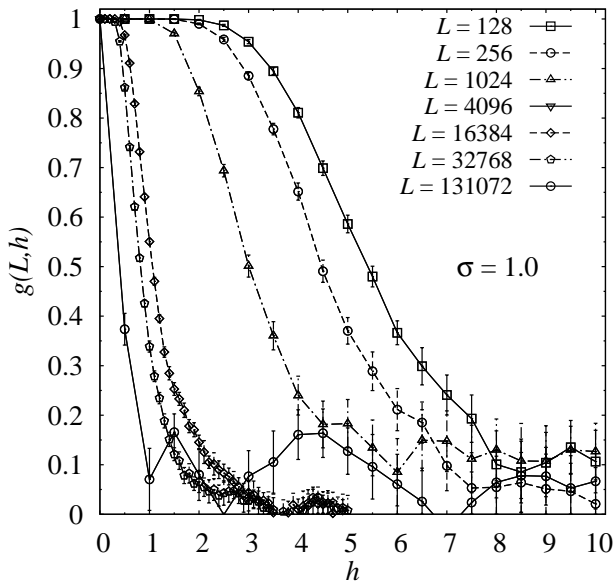}
	\caption{Binder cumulant for 
	$\sigma = 1.0$ averaged over at least $10^3$ realisations. 
	Lines are guides to the eyes only.
	\label{Binder_100}}
\end{figure}

Finally we turn to the case $\sigma=1$ where we expect no
phase transition.
Fig.\ \ref{Binder_100} shows the Binder parameter for various system sizes. 
One can see that there is no intersection between 
the curves for different system sizes, which means that $h_c = 0$. This 
is supported by the fact that the curves of the magnetization 
(not shown) for 
different system sizes do not converge towards one curve for 
$L \rightarrow \infty$, in contrast to, e.g., the case 
$\sigma = 1/4$  (cf.\ Fig.\ \ref{mag}). Thus, 
the magnetization jumps from zero for any value $h>0$
to $m=1$ for $h=0$, meaning $\beta=0$.
Nevertheless,
for the specific heat-like quantity and the susceptibilities,
we could study (not shown here) the behavior when approaching $h=0$ in the
same way as for the previously discussed values of $\sigma$.
This results in $\nu=0.40(8)$, 
$\alpha \approx 0$, $\gamma=2.19(53)$ and $\bar{\gamma}=2.51(83)$, as shown 
in Tab.\ \ref{crit_exp}.

\section{Conclusion and Outlook}
We have studied exact ground states of one-dimensional ($d=1$)
long-range random-field Ising magnets. The probability $p$ of 
placing a bond between two spins depends on the geometric distance $r$
of these spins as $p(r) \sim r^{-d-\sigma}$.
Since polynomial-time running
algorithms exist, based on a mapping to the maximum-flow problem,
we could study large systems numerically with a high
number of random samples. We studied the model
for different values of $\sigma$, which are representatives for 
the different expected behavior of the model.

Table \ref{crit_exp} summarizes 
the obtained values of the critical point and 
the critical exponents in comparison with the expected values from theory. 
In the mean-field case for $\sigma = 0.25$ the critical exponents agree 
well within error bars with the theoretical values. 
The critical point is 
consistent with values found for the Dyson hierarchical version 
\cite{Monthus11} of 
the RFIM. In the non-mean-field region for $\sigma = 0.4$, the exponents 
also agree well with theory. The critical point $h_c$ does 
not agree with the one found in reference \onlinecite{Monthus11}, but these 
points are anyway non-universal.

In the borderline case $\sigma = 0.5$, in particular the critical 
point $h_c > 0$ is an interesting result, as only statements
\cite{Bray86,Weir87,Cassandro09,Aizenman90,Aizenman89,Aizenman90_Err,
Monthus11} of the existence of a finite-disorder phase transition for 
$\sigma \neq 1/2$ have been published so far. In addition, mathematical proofs 
\cite{Aizenman90,Aizenman89,Aizenman90_Err,Cassandro09} 
do not exclude the possibility of $h_c > 0$ for $\sigma_c=1/2$ at zero 
temperature. Recent work \cite{Leuzzi13}, which was performed 
independently and in parallel to our work, support $h_c > 0$. In the 
cited work, an Imry-Ma 
argument (cf.\ also Refs.\ \onlinecite{Bray86,Weir87,Cassandro09,
Monthus11})
is given and also calculations of exact ground states were 
carried out independently of our work, 
but it was restricted to the analysis of the Binder 
cumulant and few other observables.
Nevertheless, all measured exponents agree with theory,
if one assumes the theory (cf.\ Eqs.\ \eqref{crit_exp_NMF}) for 
$1/3\le \sigma <1/2$
to be valid also at $\sigma=1/2$. Note that the value
of $\beta$ is off by a few error bars, but for values close
to zero, one would have to go to large system sizes
to see the limiting behavior.

For $\sigma = 1$, the measured critical point $h_c = 0$ 
agrees with theory as well as the value for $\beta$. Nevertheless,
the expected jumps 
\cite{Aharony83} in the magnetization as $\beta = 0$ were not observed.
As usual for first-order transitions, a
real jump can be expected to be visible only in the
thermodynamic limit, i.e.\ for huge system sizes.

The found value of the correlation length exponent $\nu$ does agree 
with theory within two error bars, where $\nu = 1/2$ is predicted 
\cite{Aharony83,Bray85} for 
$\sigma =1$. Both values for $\gamma$ and $\bar{\gamma}$ are compatible 
with the expected values if the error bars are taken into account.

Next, we check the Rushbrooke equality \cite{Fisher63} for the different 
values of $\sigma$:
\begin{equation}
	\alpha + 2 \beta + \gamma = 2.
	\label{rushbrooke}
\end{equation}
For $\sigma = 0.25$ one gets the value 
$\alpha + 2 \beta + \gamma = 2.30(54)$, which
fulfills equation \eqref{rushbrooke} within the standard error bar.
For $\sigma = 0.4$, formula \eqref{rushbrooke} yields 
$\alpha + 2 \beta + \gamma = 2.04(70)$, which is in good agreement with 
the expected value when the statistical error is taken into account. 
For the borderline case $\sigma = 0.5$ between non-mean-field region and 
the region without a non-trivial phase transition, one obtains 
$\alpha + 2 \beta + \gamma = 2.11(90)$, which fulfills equation 
\eqref{rushbrooke} within error bars. In the region, where $h_c = 0$ 
and thus $\sigma = 1$, one gets $\alpha + 2 \beta + \gamma = 2.19(53)$, 
which satisfies the scaling relation \eqref{rushbrooke} within the 
statistical error. Because of the large error bars, resulting mainly 
from the large errors of $\gamma$, the tests of the Rushbrooke equality 
are not very significant.

We compare the theoretical and estimated values of the so-called 
droplet exponent $\theta$. In the mean-field case one gets \cite{Monthus11} 
$\theta_{\text{MF}} = \gamma_{\text{MF}}/\nu_{\text{MF}}
 = 1/\nu_{\text{MF}} = \sigma$. 
For $\sigma = 0.25$, we cannot check this directly, as we have 
measured $1/\nu^*$ rather than $1/\nu$. But according to 
Eq.\ \eqref{nu_star} we get $1/\nu = 0.253 \pm 0.048$ which agrees well 
with $\theta_{\text{MF}} = \sigma$.
In the non-mean-field 
region one obtains $\theta = \bar{\gamma}/\nu - \gamma/\nu$. 
For $\sigma = 0.4$ this yields $\theta = 0.378(81)$, 
and for $\sigma = 0.5$ we get $\theta = 0.452(45)$.
In the case $\sigma = 0.4$ it agrees within one and for $\sigma = 0.5$ 
within two error bars with the prediction 
$\theta = \sigma$ by Grinstein \cite{Grinstein76}. 
But smaller deviations from this conjecture 
could not be determined as the error bars of these quantities 
are too large. In the case $\sigma = 1$, we obtain
$\theta = 0.13(16)$ which is compatible with $\theta = 0$ within the 
error bar.

The conjecture \cite{Grinstein76} $\theta = \sigma$ 
only holds for the Dyson hierarchical model \cite{Rodgers88}. 
It was shown later, that this prediction was 
pertubatively wrong at higher orders \cite{Bray86} for models with 
interaction strengths which decay like a power-law in the distance.
However, for our model, we cannot make a statement whether 
the conjecture $\theta = \sigma$ holds or not, because for 
$\sigma = 0.4,0.5$ our data does not allow the determination of 
small deviations from this conjecture because of too large error bars.

In a two exponent scenario, the Schwartz-Soffer equation 
\cite{Schwartz85} 
\begin{equation}
	\bar{\gamma} = 2 \gamma
	\label{Schwartz_Soffer}
\end{equation}
would hold. For $\sigma = 0.25$ formula \eqref{Schwartz_Soffer} is valid, 
when the statistical error is taken into account. 
In the cases $\sigma = 0.4$ and $\sigma = 0.5$, equation 
\eqref{Schwartz_Soffer} is also fulfilled within statistical errors. For 
$\sigma = 1$ the Schwartz-Soffer equation does not hold.

To summarize, the critical exponents for the investigated values 
$\sigma \in \{0.25,0.4,0.5,1\}$ 
agree well with theory, most values within one, few within two 
error bars. 
This deviation might be due to too large system sizes which are 
needed to see the infinite-size behavior. The 
Rushbrooke equality is fulfilled for all studied values of $\sigma$. 
The droplet exponent $\theta$ agrees well with theory for 
$\sigma \in \{0.4,0.5,1\}$, although a statement if the 
conjecture $\theta = \sigma$ holds is not possible.
The two-exponent scenario is supported 
by the confirmation of the Schwartz-Soffer equation for 
$\sigma \in \{0.25,0.4,0.5\}$.

For the critical case $\sigma = 1/2$, it was found that $h_c > 0$,
as for other recent numerical studies on the Dyson hierarchical
model \cite{Monthus11} and for the same diluted model \cite{Leuzzi13}
as studied here. This is an interesting result, because with 
the Imry-Ma argument \cite{Bray86,Weir87,Cassandro09,Monthus11} only 
conclusions for the cases $\sigma < 1/2$ or $\sigma > 1/2$ are possible. 
Rigorous studies \cite{Aizenman90,Aizenman89,Aizenman90_Err,Cassandro09} 
do also not exclude $\sigma = 1/2$ as possible value of a finite-disorder 
phase transition at zero temperature.
Our data allows no conclusion about the type of finite-size scaling 
behavior, as both an algebraic as well as a logarithmic behavior is 
possible.

For future studies, it could be of interest
to study the same diluted long-range model on higher dimensional lattices.
At least $d=2$ and $d=3$ should be accessible using the highly efficient
maximum-flow algorithms used here.

\begin{acknowledgments}
We would like to thank M. Moore for suggesting the project
to us. Furthermore, we thank him,
C. Monthus, A. van Enter, A. P. Young,  and T. Garel for helpful 
discussions.

The simulations were performed at the HERO cluster of the University
of Oldenburg funded by the DFG (INST 184/108-1 FUGG) and the 
ministry of Science and Culture (MWK) of the Lower Saxony State.
\end{acknowledgments}

\bibliography{RFIM.bib}

\end{document}